\documentclass[aps,showpacs,prd,preprintnumbers,nofootinbib,appendix]{revtex4}

\usepackage{bbm}                                                  

\newcommand{\nc}{\newcommand}
\nc{\beq}{\begin{equation}}  \nc{\eeq}{\end{equation}}
\nc{\bea}{\begin{eqnarray}}  \nc{\eea}{\end{eqnarray}}
\nc{\baa}{\begin{array}}     \nc{\eaa}{\end{array}}
\nc{\bit}{\begin{itemize}}   \nc{\eit}{\end{itemize}}
\nc{\ben}{\begin{enumerate}} \nc{\een}{\end{enumerate}}
\nc{\bce}{\begin{center}}    \nc{\ece}{\end{center}}
\nc{\bpm}{\begin{pmatrix}}   \nc{\epm}{\end{pmatrix}}
\nc{\bvt}{\begin{verbatim}}  \nc{\evt}{\end{verbatim}}
\def\half{\frac12}      

\def\to{\rightarrow}

\def\gesim{\,{\raise-3pt\hbox{$\sim$}}\!\!\!\!\!{\raise2pt\hbox{$>$}}\,}
\def\lesim{\,{\raise-3pt\hbox{$\sim$}}\!\!\!\!\!{\raise2pt\hbox{$<$}}\,}
\def\boldoverdot{\,{\raise6pt\hbox{\bf.}\!\!\!\!\>}}

\def\lcal{{\cal L}}

\def\vcal{{\cal V}}

\def\zBB{{\mathbbm Z}}
\usepackage{bm}
\def\tr{ \hbox{tr}}

\def\diag{\hbox{\diag}}
\def\sm{SM}

\def\gev{\hbox{GeV}}
\def\tev{\hbox{TeV}}

\def\doubleundertext#1{
{\undertext{\vphantom{y}#1}}\par\nobreak\vskip-\the\baselineskip\vskip4pt%
\undertext{\hbox to 2in{}}}
\def\inbox#1{\vbox{\hrule\hbox{\vrule\kern5pt
     \vbox{\kern5pt#1\kern5pt}\kern5pt\vrule}\hrule}}
\def\sqr#1#2{{\vcenter{\hrule height.#2pt
      \hbox{\vrule width.#2pt height#1pt \kern#1pt
         \vrule width.#2pt}
      \hrule height.#2pt}}}

\def\today{\ifcase\month\or
  January\or February\or March\or April\or May\or June\or
  July\or August\or September\or October\or November\or December\fi
  \space\number\day, \number\year}
\def\pmb#1{\setbox0=\hbox{#1}%
  \kern-.025em\copy0\kern-\wd0
  \kern.05em\copy0\kern-\wd0
  \kern-.025em\raise.0433em\box0 }
\def\up#1{^{\left( #1 \right) }}
\def\inv#1{\frac1{#1}}

\def\ui{U(1)}
\def\sumprime_#1{\setbox0=\hbox{$\scriptstyle{#1}$}
  \setbox2=\hbox{$\displaystyle{\sum}$}
  \setbox4=\hbox{${}'\mathsurround=0pt$}
  \dimen0=.5\wd0 \advance\dimen0 by-.5\wd2
  \ifdim\dimen0>0pt
  \ifdim\dimen0>\wd4 \kern\wd4 \else\kern\dimen0\fi\fi
\mathop{{\sum}'}_{\kern-\wd4 #1}}
\usepackage{color}

\def\lsp{\qquad}

\def\ab{X}
\def\nab{G}

\nc{\scs}{S_{\hbox{\tiny CS}}}
\nc{\acs}{\alpha_{\hbox{\tiny CS}}}
\nc{\lcs}{\Lambda_{\hbox{\tiny CS}}}
\nc{\la}{\lambda}
\nc{\ep}{\epsilon}
\nc{\be}{\beta}
\nc{\al}{\alpha}
\nc{\de}{\delta}
\def\lcal{{\cal L}}
\def\tr{{\rm Tr}}

\begin{document}

\preprint{IFT-07-02 \cr UCRHEP-T434}

\title{Note on the strong CP problem from a 5-dimensional perspective\\
{\it -- the gauge-axion unification --}}

\author{Bohdan GRZADKOWSKI}
\email{bohdan.grzadkowski@fuw.edu.pl}
\affiliation{Institute of Theoretical Physics,  University of Warsaw,
Ho\.za 69, PL-00-681 Warsaw, Poland}

\author{Jos\'e WUDKA}
\email{jose.wudka@ucr.edu}
\affiliation{Department of Physics, University of California,
Riverside CA 92521-0413, USA}

\begin{abstract}
We consider 5 dimensional gauge theories where the 5th direction 
is compactified on an interval. The Chern-Simons (CS) terms (favored by
the naive dimensional analysis) are discussed. A simple scenario with 
an extra $U(1)_X$ gauge field that couples to $SU(3)_{\rm color}$ through a CS term
in the bulk is constructed. 
The extra component of the Abelian gauge field plays a role of the axion (gauge-axion unification), which
in the standard manner solves the strong CP problem easily avoiding most of  
experimental constraints. Possibility of discovering the gauge-unification at the LCH is discussed.

\end{abstract}


\maketitle

\section{Introduction}

In the Standard Model (SM), the Higgs mechanism is responsible for generating
fermion and vector-boson masses. 
Although the model is renormalizable
and unitary, it has severe naturality problems
associated with the so-called ``hierarchy problem''.
At loop-level this problem reduces to the fact 
that the quadratic corrections tend to 
increase the Higgs boson mass up to the UV cutoff of the theory.
Extra dimensional extensions of the SM offer a novel
approach to gauge symmetry breaking in which the hierarchy problem could be
either solved or at least reformulated in terms of the geometry of the higher-dimensional space.

Other inherent problems of the SM could also be addressed in
extra-dimensional scenarios. For instance, within the SM
the amount of CP violation is not sufficient to explain the observed baryon 
asymmetry~\cite{Barr:1979ye}, the gauge-Higgs unification scenario offers a possible solution
since in such models the geometry can be a new source
of explicit and spontaneous CP violation~\cite{Grzadkowski:2004jv}. 
In this note we shall prove that the strong CP problem could be solved introducing
an appropriate Chern-Simons (CS) terms in 5D~\footnote{For other attempts to solve the strong CP problem by 
5 dimensions see \cite{Aldazabal:2002py}.}. The scenario leads to an attractive possibility of gauge-axion unification.

\section{Hierarchy of effective operators}

We will first consider models in $ D = 5$ dimensions with fermions, gauge bosons and
scalars propagating throughout the $D$-dimensional bulk, and 
some unspecified matter localized on lower dimensional manifolds (branes). 
Though these models are non-renormalizable
it is possible to define a hierarchy of possible terms in the
Lagrangian that allows for a proper perturbative expansion; the procedure
is a simple application of the arguments used in the naive dimensional 
analysis (NDA) \cite{Manohar:1983md}, see the Appendix.
This hierarchy is specified by assigning  to each gauge invariant
operator an index $s=d_c+b'+(3f/2)-4$, ($d_c$ is the number of covariant
derivatives, $f$ and $b'$ the number of fermion and scalar fields).
As it is shown in the Appendix the least suppressed operators are those that have the index $s=0$:
\beq
F^2; \quad \bar\psi D \psi; \quad |D\phi|^2; \quad \bar\psi\phi\psi ; \quad \phi^4\,,
\label{eq:index_zero}
\eeq
where $F$ denotes the generic gauge tensor,
$ \phi $ a generic scalar, and $ \psi$ generic fermions.

The $s=1$ operators not containing scalar fields 
are ($A$ denotes a generic gauge field)
\beq
A F^2 ; \quad \bar\psi F \psi\,,
\label{eq:index_one}
\eeq
whose coefficients are naturally suppressed by  $ 1/(24\pi^3) $,
together with all brane terms, presumably including the \sm\ Lagrangian multiplied 
by
$ l_4^{-1} \delta(y-y_o) $.
The first operator in (\ref{eq:index_one}) corresponds to the 5-dimensional Chern-Simons 
(CS) term, while the second includes all magnetic-type couplings.
Operators of index $s=1$ containing $ \phi $ are of the form $ D^4 \phi $, $ D^2 \phi^3 $,
or $ D \bar\psi \psi \phi $. 

The NDA argument favors the presence of a CS term (if only 5D vector bosons are present the CS term is the only
bulk operator with index $s=1$),
with a coefficient as large as $ 1/(24\pi^3) $.
Of course, it is still possible that there exist additional  symmetries that forbid this
term, however if present, the CS term can generate interesting effects.

Hereafter we shall consider a 5D model containing $U(1)_X$ and
$SU(3)_{\rm color}$ bulk gauge fields,
denoted by $X$ and $G$ respectively.
Application of the NDA for this case (where there are no bulk fermions) yields the following
action up to index $s=1$
\bea
S &=& \int_{X^5} d^5 x \left\{-\inv4 \ab_{MN}\ab^{MN}  -\inv2 \tr  \left[ \nab_{MN}\nab^{MN}\right]
+ \right.\cr
&& - \left. \inv{24\pi^3}
\epsilon^{LMNPQ} \left[
 c_1 g_5^{'} g_5^2 X_L \tr \left(\nab_{MN} \nab_{PQ} \right) 
+ c_2 g_5^{' 3} X_L \ab_{MN} \ab_{PQ} + \right.\right. \cr
&& + \left.\left. c_3 g_5^3  
\tr \left(G_L \nab_{MN}\nab_{PQ} + \frac{i}{2} G_L G_M G_N \nab_{PQ} - \frac1{10}G_L G_M G_N G_P G_Q \right)
 \right]
 \right\} 
 + \inv{16\pi^2} S_{\rm brane} \,
\label{eq:full-action}
\eea
where $\ab_{MN}$ and $\nab_{MN}$ are, respectively, the field strength tensors 
for the Abelian and non-Abelian groups~\footnote{The convention for the 
antisymmetric tensors which we
follow is such that $\epsilon_{01234}=\epsilon_{0123}=1$ for the metric tensor
$\eta_{MN}={\rm diag}(1,-1,-1,-1,-1)$ and $\eta_{\mu\nu}={\rm diag}(1,-1,-1,-1)$.
We assume that the non-Abelian group generators, $T^a$ are Hermitian and 
normalized according to $ \tr T^a T^b = 2^{-1} \delta_{ab} $.} with the 5D gauge couplings
respectively denoted by $g_5^{'}$ and $g_5$; $ c_{1,2,3} $ are undetermined
numerical  constants, presumably of $O(1) $.
In our specific applications we will consider models constructed
on the space-time $X^5=M^4\times [0,R]$, and we 
will concentrate on the ``mixed'' Chern-Simons term proportional
to $g_5^{'} g_5^2$. We will assume that all \sm\ fields are neutral under $ \ui_X $.
Hereafter, whenever possible, in order to make the analysis as model independent as possible,
we will avoid referring to any details of the embedding of the SM into 5D. The only assumption we make
is that the SM is localized on one or perhaps both ends of the interval $[0,R]$.

\section{Solving the  strong CP problem from a 5D perspective}

As shown above, the NDA favors the CS term as an operator of index $s=1$.  
We will argue that the presence of this term allows for a simple solution
to the strong CP problem.

As it is well known, 
in a basis where the Yukawa matrices are diagonal,
the phases of the Kobayashi-Maskawa matrix are responsible for
all electroweak CP violation effects. There is, however,
an additional (``strong'') CP-violating term 
allowed by the symmetries of the 4D SM Lagrangian:
\beq
\lcal_{\rm QCD~CP} = \theta \frac{\alpha_s}{16\pi} 
{\rm Tr}\left(\nab_{\mu\nu}\tilde \nab^{\mu\nu}\right)\,,
\label{theta_SM}
\eeq
where $\nab_{\mu\nu}$ is the QCD field strength tensor,
$\tilde \nab^{\mu\nu}=\epsilon^{\mu\nu\alpha\beta}\nab_{\alpha\beta}/2$,
and $\alpha_s \equiv g^2/(4 \pi)$ for $g$ the SM 4D QCD gauge coupling constant.
In the process of diagonalizing the Yukawa matrices, quark fields undergo
a chiral rotation, which generates the same structure as in (\ref{theta_SM})
(within the path-integral formulation this results from a non-trivial 
Jacobian for the fermionic measure \cite{Fujikawa:1979ay}); therefore 
the total effect of the strong CP violation is 
parameterized by the effective coefficient 
$\theta_{\rm eff}\equiv \theta+\theta_{\rm weak}$. The experimental 
data (EDM of the neutron) indicates that $ |\theta_{\rm eff}| \lesim 10^{-9} $  
\cite{Yao:2006px}; this is referred to as the strong-CP ``problem''
since none of the symmetries of the \sm\ requires such a strong suppression.

Models in extra dimensions offer new possibilities to solve this problem
due to a possibility of constructing the Chern-Simons terms. Specifically, we will assume that the color
gauge fields $G_N^a$ propagate in the bulk, but that the rest of the SM
fields are confined to one or two branes located at $ y =0 $ and $ y=R $.
In addition we assume the presence of an
Abelian gauge field $ X_N$ also propagating in the bulk.
For the 5D models being considered here, the 
QCD strong-CP term (\ref{theta_SM}) can be written as follows:
\beq
S_{\rm brane} = \frac{\alpha_s}{16\pi^2}\int d^5x \left[
\theta_L\delta(y) + \theta_R\delta(y-R)\right]
\tr\left(\nab_{\mu\nu}\tilde{\nab}^{\mu\nu}\right)\,,
\label{scpv}
\eeq
where $\theta_{R,L}$ are constant parameters.

Among the various terms in (\ref{eq:full-action}) we will concentrate on the effects of
the mixed CS term:
\beq
\scs=
-\frac{g_5^{'}g_5^2 c_1}{24\pi^3}\;\int_{X^5} d^4x \; dy\; \epsilon^{LMNPQ} X_L 
\tr \left( \nab_{MN} \nab_{PQ}
 \right)\, .
\label{cs-b}
\eeq
The action (\ref{cs-b}) is not automatically gauge invariant under the $U(1)_X$. 
However, using the Bianchi identity
$\ep^{NMQPR} D_Q \nab_{PR} =0$, 
one can show that under the Abelian transformation
\beq
X_L \to X_L^\prime=X_L+\partial_L \lambda_X
\eeq
the change in $\scs$ is localized on the boundary of the 
space~\footnote{This assumes that $\la_X$ is not a constant.}.
\beq
\de \scs=  \left.
\frac{g_5^{'}g_5^2 c_1}{24\pi^3}
\;\int_{M^4}d^4x \;\la_X \;\epsilon^{\mu\nu\al\be} {\rm Tr}\left( \nab_{\mu\nu}
\nab_{\al\be}\right)\right|^{y=R}_{y=0}
\label{cs-var}
\eeq
There are various ways of insuring that this vanishes. One can, for example,
add an appropriate set of chiral fermions on the two branes; in this case
the anomaly generated by these fermions can be adjusted so that it cancels
(\ref{cs-var}), see e.g. \cite{Hill:2006ei}. Brane scalars
can be also arranged to have the same effect \cite{Aldazabal:2002py}, \cite{Hill:2006ei}
provided they couple to $\ep^{\mu\nu\al\be}\tr(\nab_{\mu\nu} \nab_{\al\be})$.
A simpler alternative, which we will adopt here, is  to impose appropriate boundary
conditions such as $\la_X\tr(\nab^2)|_{y=0}=\la_X\tr(\nab^2)|_{y=L}$.

Variation of the total  action (\ref{eq:full-action}) with $ c_2= c_3 =0 $
and $c_1=1$ leads to the following equations of motion for the gauge fields:
\beq
D_B \nab^{BA} = J^A +{\rm brane~terms}
\qquad {\rm and} \qquad
\partial_B \ab^{BA} = j^A  +{\rm brane~terms}  \,,
\label{em}
\eeq
with the following Chern-Simons currents
\beq
J^A = \frac{g_5^{'}g_5^2}{24 \pi^3}  \epsilon^{ABCDE} \ab_{BC}\nab_{DE};
\lsp
j^A = \frac{g_5^{'}g_5^2}{24 \pi^3}\epsilon^{ABCDE} \tr\left(\nab_{BC}\nab_{DE}\right)
\label{cur}
\eeq
The brane terms in (\ref{em}) originate from possible couplings of the
bulk gauge fields to the fields localized on the branes.

For the extremum of the action the following boundary conditions (BC)
must be fulfilled: 
\beq
\left. {\rm tr}\left[
\left( \nab_{4 \mu} - \frac{g_5^{'}g_5^2}{6 \pi^3}  X^\nu
\tilde \nab_{\mu\nu} \right) 
\delta G^\mu \right]  \right|_{y=0}^{y=R}=0 \qquad {\rm and} \qquad
\left.\ab^{4 \mu} \delta X_\mu \right|_{y=0}^{y=R}=0
\label{bc_two}
\eeq
Here we will restrict ourselves to theories containing massless zero-modes (gluons)
of the non-Abelian gauge field. This implies 
a unique choice of BC for $SU(3)_{\rm color}$: 
\beq
\partial_yG_\mu^a|_{y=0,R}=0 , \qquad G_4^a|_{y=0,R}= 0.\,;
\label{bcA}
\eeq
these conditions imply $\nab^a_{4\mu}|_{y=0,R}=0$. For the Abelian field
we require
\beq
X_\mu|_{y=0,R}=0, \qquad  \partial_y X_4|_{y=0,R}=0 \,,
\label{Bbc}
\eeq
so that  $\ab_{\mu\nu}|_{y=0,R}=0$. It follows that
the BC (\ref{bc_two}) are satisfied~\footnote{We thank Kin-ya Oda for a discussion at this point.}. 

The resulting Kaluza-Klein (KK) expansions read
\beq
\baa{ll}
G_\mu^a(x,y) =  R^{-1/2} \sum_{n=0} d_n G_\mu^{a\ (n)}(x)\cos m_ny \qquad
& G_4^a(x,y) = R^{-1/2}  \sqrt{2} \sum_{n=1} G_4^{a\ (n)}(x) \sin m_ny \\
X_\mu(x,y) = R^{-1/2}  \sqrt{2} \sum_{n=1} X_\mu^{(n)}(x) \sin m_ny \qquad
& X_4(x,y)  = R^{-1/2} \sum_{n=0} d_n X_4^{(n)}(x) \cos m_ny 
\eaa
\eeq
where $m_n=\pi n/R$ and $d_n = 2^{(1-\delta_{n,0})/2}$. 
The zero-mode $G_\mu^{a\ (0)}(x)$ is the standard 4D gluon;
it is also clear that the model also contains a massless 4D scalar $X_4^{(0)}(x)$.

Let's focus now on the Abelian gauge transformations. In order to preserve the BC, the gauge 
function $\lambda_X(x,y)$ must satisfy the following constraints: 
\beq
\qquad \partial_\mu \lambda_X|_{y=0,R}=0, \qquad \partial^2_y\lambda_X|_{y=0,R}=0 
\eeq
That implies a corresponding KK expansion for the Abelian gauge function
\beq
\qquad \lambda_X(x,y)  =\sum_{n=1}\lambda_X^{(n)}(x)\sin m_n y + \beta y \label{Akklam}
\eeq
where $\beta$ is a constant. The 4D vector and scalar fields transform as  
\beq
X_\mu^{(n)} \to X_\mu^{(n)}+\frac{1}{\sqrt{2}}\partial_\mu \lambda_X^{(n)}
\qquad
X_4^{(n)}\to \left\{
\baa{ll} 
X_4^{(0)}+\beta &\qquad {\rm for}\qquad n=0\\
X_4^{(n)}+\frac{m_n}{\sqrt{2}} \lambda_X^{(n)} &\qquad {\rm for}\qquad n>0
\eaa
\right.
\,.
\label{gtrans}
\eeq
In the following we will take $\beta=0$, which is the simplest condition 
ensuring the  gauge symmetry of the CS action~\footnote{
This is also a natural choice for $S^1/Z_2$ orbifold models
since it insures that $X_\mu(x,-y)=-X_\mu(x,y)$, $X_4(x,-y)=X_4(x,y)$
and $X_N(x,y+2R)=X_N(x,y)$ are preserved under gauge transformations.}.

In order to discuss phenomenological predictions of the model let us expand the CS action into KK modes:
\beq
S_{\rm CS} = \frac{R}{12\pi^3} \frac{g_5^{'}}{R^{1/2}}
\frac{g_5^2}{R} c_1 \int d^4 x
\Biggl[ X_4 \up 0 \; \tr G_{\mu\nu}\up0 \tilde G^{\mu\nu}{}\up0
+ 2 \partial_\mu X_4 \up 0 \; \sum_{n=1}^\infty \tr G_\nu\up n D_\rho G\up n_\sigma \epsilon^{\mu \nu \rho  \sigma}
- 4 \tr \tilde G^{\mu\nu}{}\up0 \sum_{n=1}^\infty \Theta_{\mu\nu}\up n + \cdots \Biggr]
\label{cs_strong}
\eeq
where
\beq
D_\mu \equiv \partial_\mu + i g \left[ G_\mu\up 0 , \cdot \right]  \quad
G_{\mu\nu}\up 0 \equiv \partial_\mu G_\nu \up 0 - \partial_\nu G_\mu\up 0 + i g
\left[G_\mu\up 0, G_\nu\up 0 \right] 
\label{der}
\eeq
for $g = g_5/\sqrt{R}$ and
\beq
\Theta_{\mu\nu}\up n \equiv 
\half \Biggl[ \left( \partial_\mu X_4\up n G_\nu\up n - \partial_\nu X_4\up n G_\mu\up n \right)
- \left( \partial_\mu X_\nu\up n G_4\up n - \partial_\nu X_\mu \up n G_4\up n \right) 
- m_n \left( X_\mu\up n G_\nu\up n - X_\nu\up n G_\mu\up n \right) \Biggr]
\label{theta}
\eeq
Ellipsis in (\ref{cs_strong}) stands for terms (irrelevant for any practical applications) that involve
four non-zero KK modes. Expanding the kinetic terms of (\ref{eq:full-action}), one can verify that indeed
$\nab_{\mu\nu}^{(0)}$ corresponds to the SM QCD gluon
(which is present due to our having adopted (\ref{bcA})), 
while $X_4^{(0)}(x) = a(x)$ can play the role of the axion. 
The lowest-order terms conform the
usual QCD action, the axion kinetic term and the axion-gluon interactions~\footnote{It turns out
that each term in the KK expansion of (\ref{scpv}) is a total derivatives (as they emerge form the full derivative
$\tr[\nab_{\mu\nu}\tilde{\nab}^{\mu\nu}]$). Only the zero-mode contribution will be relevant as it contributes
to the effective non-perturbative axion potential, other terms could be dropped.}:
\beq
S_{\rm low}\up 0 =  \int_{M^4}\left\{
- \frac{1}{2}\tr\left(\nab_{\mu\nu}\nab^{\mu\nu}\right) +
\frac12 \partial_\mu a\partial^\mu a +
\frac{\alpha_s}{16\pi}\left(\frac{a}{f_a}+\theta_{\rm eff}\right)
{\rm Tr}\left(\nab_{\mu\nu}\tilde \nab^{\;\mu\nu}\right)
\right\} \,,
\label{len}
\eeq
where $\theta_{\rm eff}\equiv \theta_L+\theta_R$
and we dropped the $(0)$ superscript in $\nab$.
Adopting the NDA estimation of the CS coefficient
one obtains for the axion decay constant
\beq
f_a^{-1}=\frac{16 g^{'}}{3 \pi}R
\label{fnda}
\eeq
where $g^{'}$ is the 4D Abelian gauge coupling, $g'=g_5'/\sqrt{R}$,
and $\alpha_s = g^2/(4 \pi)$.
Note that for this mechanism of axion generation to work, the extra Abelian gauge symmetry must be broken
by the boundary conditions (Scherk--Schwarz breaking) so that no additional 
massless vector boson associated with  $X_\mu$ is present. The only low-energy remnant
of $X_M$ is the axion $a(x)$. The crucial advantage of the model presented here is the unification of the axion and the $U(1)$
5D gauge field. There are serious attempts to construct in 5D a realistic gauge-unification theory \cite{Hosotani:2006qp}.
Those models combined with the scenario discussed here could provide an interesting alternative for a theory of electroweak interactions that 
offers the scalar sector of 4D theory fully unified with a gauge fields (solving the hierarchy problem \cite{Hosotani:2006qp} and
the strong CP problem at the same time). As it will be discussed below the gauge-axion unification is consistent with
the existing experimental constraints and there is a chance to test the scenario at the LHC.

As in the standard Peccei-Quinn scenario the effective axion coupling
$(a/f_a+\theta_{\rm eff})$
relaxes to zero through  instanton effects, solving the strong CP problem 
dynamically.
The axion mass is generated in a standard manner~\cite{Cheng:1987gp} 
\beq
m_a=\frac{f_\pi m_\pi}{f_a}\frac{\sqrt{m_u m_d}}{m_u+m_d}=0.6{~\rm eV}
\frac{10^7{\rm ~GeV}}{f_a}\,,
\label{mafa}
\eeq
and no strictly massless scalars remain in the spectrum.

Let us discuss consequences of the remaining interactions in the 5D CS term (\ref{cs-b}) that consists 
of terms quadratic and quartic in the non-zero KK modes. We will
focus (for obvious phenomenological reasons) on the quadratic terms 
shown explicitly in (\ref{cs_strong}). Of course, there are other terms involving the
heavy fields generated by the kinetic part of the action (\ref{eq:full-action}), those have
have been considered previously in the literature, see e.g.~\cite{Dicus:2000hm}~.

Because of its relatively large coupling,
the very last term ($\propto m_n$)  in  (\ref{cs_strong}),
will produce the most noticeable effects at the LHC.
Therefore let us consider the production of heavy gluons $ G_\mu\up n$
and vector bosons $ X_\mu\up n $ (with $ n \ge 1$) at the LHC.
At the partonic level
the leading contributions are the following: $G G \to G^\star \to G \up n X \up n$ 
and $G  G  \to G \up n X \up n$. 
Since the SM fields do not carry $U(1)_X$ quantum numbers, the $X_\mu\up n$ bosons are stable
at the tree level; on the other hand, heavy gluons  $G_\mu \up n$ couple to SM quarks located on a brane.
Therefore the experimental signature for the above reactions would be missing energy 
and momentum (carried away by the stable $X_\mu \up n$) and two jets from the $G_\mu \up n$ decays. 
Let us compare the amplitude strength for this 
process with the standard QCD two jet production amplitude.
Adopting the estimate of the CS coupling from the NDA in  (\ref{cs_strong})
we find that the ratio of the $X_\mu \up n G_\nu \up n G_\alpha $  coupling
to the SM triple gluon vertex is of the order of
\beq
\frac{g'}{g}\frac{\alpha_s}{3\pi}\; n \sim \frac{g'}{g}10^{-2}\; n
\eeq
Since $n\sim 1$ (otherwise KK modes are too heavy to be produced), it seems that it may be difficult 
to detect $G \up n X \up n$ over the two-jet QCD background. 
Nevertheless it should be noticed, that the huge amount ($\sim \tev$) of missing energy (carried away by the stable and heavy 
$X_\mu \up n$) may enhance the signal relative to
the QCD background very efficiently, and that the large gluon luminosity of the LHC could be sufficient to provide
enough events to test the scenario. Though these expectations are supported by
the results for similar processes at the Tevatron ~\cite{Tevatron},
a dedicated Monte Carlo study would be needed to resolve this issue definitively;
this, however, lies beyond the scope of this note.

Other possible signature of the axion being the 4th component of 5D gauge field 
could be the heavy gluon production process
through a virtual axion exchange: $G G \to a^\star \to G \up n G \up n$ for $n \ge 1$. 
The amplitude for this process is generated by
the first two terms in  (\ref{cs_strong}). It is straightforward to find that the order of magnitude for the amplitude
normalized to two gluon ($G G$) production is the following:
\beq
\frac{\alpha^\prime}{9 \pi^2}\alpha_s \; n^2 \sim 10^{-3} \alpha^\prime\; n^2\,,
\eeq
where $\alpha^\prime \equiv g^{\prime 2}/(4 \pi)$. If $\alpha^\prime \sim \alpha_s$ then for small $n$ the amplitude is suppressed
by the factor $10^{-4}$. Since both  $G \up n G \up n$ and $G G $ states decay roughly the same way 
(the signature is $n \geq 4$ jets in the final state), it would be a real challenge to see the axion exchange 
over the standard QCD background~\footnote{Note also that the amplitude
receives contributions from the other terms in the action.}. 

Let us assume that the axion mass $m_a$ (or equivalently the decay constant $f_a$) is known. Then the definite test of the
model discussed here would be a verification of the gauge-axion unification that is caused by the fact that the axion is 
a component of the 5D gauge field $X_M$. The important consequence of the unification 
is that the total cross section for $G \up n X \up n$ production is predicted including 
the normalization. Therefore the measurement of $\sigma_{\rm tot}(G \up n X \up n)$ shall provide the definite experimental test of the model.

Concluding the review of various possible experimental tests of gauge-axion unification discussed here, one can say
that, because of a hudge missing energy ($\sim \tev$), the process $G G \to G \up n X \up n$ provides the cleanest signature, that makes the observation of the signal plausible.

For the model being considered here the axion decay constant $f_a$ is determined by the geometrical scale
 $R^{-1}$ (if the NDA arguments are applied), 
therefore experimental limits on $f_a$  constrain the size of the compact dimension dimension.
However, it should be emphasized that most of these constraints rely
on effects produced by the coupling of the axion to two photons, and this
coupling is {\em absent} in our model (to leading order). (For a review of experimental constraints see \cite{Yao:2006px}.)
Nevertheless there exists  a bound that
should be obeyed also by our photofobic axion; this is the so called ``misalignment''
lower axion mass
limit that originates from the requirement that the contribution to the
cosmic critical density from the relaxation of the axion field 
($\theta_{\rm eff}\to 0$) does not overclose the universe.
The resulting constraint~\cite{Yao:2006px}, $m_a > 10^{-6}$~eV, leads to
 $R^{-1} \lesim  10^{13}\gev$, having used
(\ref{fnda}-\ref{mafa}) and taken $g^{'} = {\cal O}(1)$. Note that the NDA estimate of the CS coupling
was crucial to derive the limit on $R$.

\section{Conclusions}

We shown that an extension of naive dimensional analysis to 
5D gauge theories naturally allows relatively large coefficients in front of Chern-Simons (CS) terms.
The strong CP problem was discussed within a simple scenario containing 
a new $U(1)_X$ gauge field and the $SU(3)_{\rm color}$ 
gauge fields propagating in the bulk, and interacting through a
a mixed CS term. 
Adopting appropriate boundary conditions, the CS term
was shown to be gauge invariant (without any need for brane matter).
The zero mode of the extra component of the new Abelian gauge field was seen to play a role of the axion
(gauge-axion unification), which in the standard manner receives the instanton-induced potential, so that the strong CP problem
(localized on the branes) disappears while the axion receives a mass. 
In the effective low-energy regime, the axion couples
only to gluons, therefore most of the limits on the axion decay constant do not apply 
in the context of this model. It was shown that the most promising test of the gauge-axion
unification is the process of $G \up n X \up n$ production: $G G \to G \up n X \up n$.
The hudge missing energy ($\sim \tev$) carried away by the stable and heavy 
$X_\mu \up n$ is believed to provide a sufficiently clean signature of the final state.

\begin{center}
{\bf APPENDIX}
\end{center}
In this appendix we provide, for completeness, a summary of the application of
Naive Dimensional Analysis (NDA) to higher-dimensional models.
The NDA allows to determine 
the scale $\Lambda$ at which the theory becomes strongly interacting.
For that purpose let us compare two graphs with the same number of external legs, one
of which has an additional gauge-boson propagator. This second graph will be
suppressed with respect to the first by the factor
\beq
\Lambda^\delta g^2 l_{4 + \delta}^{-1} ; \quad l_D =(4 \pi)^{D/2} \; \Gamma(D/2) \,,
\eeq
where $g$ denotes the gauge coupling constant, and
$ l_D $ is the geometric loop factor obtained form integrating over
momentum directions
(note that in $D = 4 + \delta $ dimensions $g$ has a mass 
dimension of $-\delta/2$).
For a strongly interacting theory we impose the NDA requirement that
the loop corrections be of the same order as the lowest-order value;
this requires
\beq
\Lambda \sim \left( l_{4 + \delta} g^{-2} \right)^{1/\delta}\,.
\label{cut}
\eeq
The same NDA requirement allows an estimate of the coefficients in front 
of effective operators. For this we consider a generic vertex of the form
\beq
{\cal V}=\lambda \; \Lambda^D (2\pi)^D\delta^D(\sum p_i)
\left(\frac{g\, \psi}{\Lambda_\psi^{3/2}}\right)^f
\left(\frac{p}{\Lambda}\right)^d
\left(\frac{g\,A_M}{\Lambda}\right)^b
\left(\frac{g\, \phi}{\Lambda_\phi}\right)^{b'}\,;
\label{oper}
\eeq
where scale appropriate for the vector fields and derivatives (they enter together
through the covariant derivative) was chosen to be $\Lambda$, while the coefficient
$\lambda$, the fermionic scale ($\Lambda_\psi$) and the scalar scale ($\Lambda_\phi$)
are to be determined.
The requirement to reproduce the starting operator by radiative corrections 
determines the maximal value of $\lambda$ and minimal scales $\Lambda_\psi$, $\Lambda_\phi$
that are allowed by perturbativity
\beq
\lambda=l_{4+\delta}^{-1} \qquad {\rm and} \qquad 
\Lambda_\psi=\Lambda_\phi =\Lambda.
\label{limits}
\eeq
Let us now restrict ourselves to 5d theories, $\delta=1$, and define the 
``index'' of a vertex by 
\beq
 s = d_c + b' + \frac32 f - 4 ; \quad d_c = d+ b\,.
\label{eq:index.def}
\eeq
where $d_c$ is the number of covariant derivatives
present in the vertex $ \vcal $.
If an $L$-loop graph contains $V_n$ vertices with indices $s_n$,
then the vertex corresponding 
to this graph has an index 
\beq
s = L + \sum_n  V_n s_n. 
\label{eq:hierarchy}
\eeq
In terms of $s$
the coefficient of a given operator is
(see also \cite{Georgi:1992dw})
\beq
\left(\frac1{24 \pi^3}\right)^s \times ({\rm 
the~powers~of~}g{\rm~needed~to~get~a~dimension~5~object})\,;
\label{eq:prescription}
\eeq
and $ \Lambda = 24\pi^3/g^2$.

If the indices of all vertices are non-negative, then it follows from (\ref{eq:hierarchy})
that $ s \ge s_n $ for all $n$.
This implies that if $ \vcal $ has index $s$, then
only operators with indices $ \le s $ can renormalize the coefficient of $ \vcal $ 
and
we can then define a hierarchy according to the value of $s$ in the sense that 
we can consistently assume that operators with higher indices are generated
only by higher orders in the loop expansion .
This would be spoiled if the theory has vertices with negative indices, 
(as an addition of an internal line attached by vertices
with $s_n<0$ decreases $s$, so an extra loop leads to less suppressed operator) 
which corresponds to the case
$ d_c =f =0 ,~b'=3$, according to the definition (\ref{eq:index.def}).
In order to define a hierarchy one should accordingly require that all
terms cubic in the scalar fields be absent~\footnote{This statement holds within NDA, of course,
if the coefficients of super-renormalizable operators are tuned to be small, then their effects are suppressed so that the problem of consistency disappears.} due to an additional symmetry such as a discrete $\zBB_2 $
under which the $ \phi $ are odd, by gauge invariance,
(as in the \sm) or just by an absence of scalar fields (as in this note where we are considering 
only vector bosons in 5D therefore the cubic scalar interactions 
cannot be constructed and the hierarchy of operators
is given just by (\ref{eq:prescription}) without any other constraints). 
Fermion fields are assumed to transform appropriately under
this symmetry, so as to allow all desirable scalar-fermion couplings.

In order to include consistently possible brane terms 
in the hierarchy we note that this type of interactions are naturally generated
by the bulk terms in a compactified space at the one loop level~\cite{Georgi:2000ks}. ~
It is then natural add $1$ to $s$
whenever a localizing factor of the form $ \delta(y-y_o) $ is present. In addition the
geometric suppression factor for these terms equals $ l_4 = 16\pi^2 $
that replaces $ l_5 = 24\pi^3 $ present in (\ref{eq:prescription}); see also \cite{Chacko:1999hg}.

\acknowledgments
This work is supported in part by the Ministry of Science and Higher Education 
(Poland) in years 2006-8 as research project N202~176~31/3844, 
by EU Marie Curie Research Training Network HEPTOOLS,
under contract MRTN-CT-2006-035505 and
by the U.S. Department of Energy grant No.~DE-FG03-94ER40837.
B.G. acknowledges the support of the European Community under MTKD-CT-2005-029466
Project, he also
thanks Jacek Pawelczyk and Kin-ya Oda 
for their interest at the beginning of this project,
and Zygmunt Lalak for being a patient witness of his
struggle while this work was emerging.


\begin{thebibliography}{99}

\bibitem{Barr:1979ye}
S.~M.~Barr, G.~Segre and H.~A.~Weldon,
Phys.\ Rev.\ D {\bf 20}, 2494 (1979).

\bibitem{Grzadkowski:2004jv}
  B.~Grzadkowski and J.~Wudka,
  Phys.\ Rev.\ Lett.\  {\bf 93}, 211603 (2004)
  [arXiv:hep-ph/0401232];
  Phys.\ Rev.\  D {\bf 72}, 125012 (2005)
  [arXiv:hep-ph/0501238];
  Acta Phys.\ Polon.\  B {\bf 36}, 3523 (2005)
  [arXiv:hep-ph/0511139].
   
\bibitem{Aldazabal:2002py}
  G.~Aldazabal, L.~E.~Ibanez and A.~M.~Uranga,
  JHEP {\bf 0403}, 065 (2004)
  [arXiv:hep-ph/0205250];
%
  K.~I.~Izawa, T.~Watari and T.~Yanagida,
  Phys.\ Lett.\ B {\bf 534}, 93 (2002)
  [arXiv:hep-ph/0202171];
%
  K.~W.~Choi,
  Phys.\ Rev.\ Lett.\  {\bf 92}, 101602 (2004)
  [arXiv:hep-ph/0308024];
%
  A.~Fukunaga and K.~I.~Izawa,
  Phys.\ Lett.\ B {\bf 562}, 251 (2003)
  [arXiv:hep-ph/0301273];
%
  R.~Harnik, G.~Perez, M.~D.~Schwartz and Y.~Shirman,
  JHEP {\bf 0503}, 068 (2005)
  [arXiv:hep-ph/0411132].
   
\bibitem{Manohar:1983md}
  A.~Manohar and H.~Georgi,
  Nucl.\ Phys.\  B {\bf 234}, 189 (1984).
 
\bibitem{Fujikawa:1979ay}
  K.~Fujikawa,
  Phys.\ Rev.\ Lett.\  {\bf 42}, 1195 (1979);
  Phys.\ Rev.\  D {\bf 21}, 2848 (1980)
  [Erratum-ibid.\  D {\bf 22}, 1499 (1980)].

\bibitem{Yao:2006px}
  W.~M.~Yao {\it et al.}  [Particle Data Group],
  J.\ Phys.\ G {\bf 33} (2006) 1.

\bibitem{Hill:2006ei}
C.~T.~Hill,
Phys.\ Rev.\ D {\bf 73}, 085001 (2006)
[arXiv:hep-th/0601154].

\bibitem{Hosotani:2006qp}
  Y.~Hosotani, S.~Noda, Y.~Sakamura and S.~Shimasaki,
  Phys.\ Rev.\  D {\bf 73}, 096006 (2006)
  [arXiv:hep-ph/0601241].
  
\bibitem{Cheng:1987gp}
H.~Y.~Cheng,
Phys.\ Rept.\  {\bf 158}, 1 (1988).

\bibitem{Dicus:2000hm}
  D.~A.~Dicus, C.~D.~McMullen and S.~Nandi,
  Phys.\ Rev.\  D {\bf 65}, 076007 (2002)
  [arXiv:hep-ph/0012259];
%
  A.~Muck, A.~Pilaftsis and R.~Ruckl,
  Phys.\ Rev.\  D {\bf 65}, 085037 (2002)
  [arXiv:hep-ph/0110391].
  
\bibitem{Tevatron}
  T.~Han, D.~L.~Rainwater and D.~Zeppenfeld,
  Phys.\ Lett.\  B {\bf 463}, 93 (1999)
  [arXiv:hep-ph/9905423];
  A.~Abulencia {\it et al.}  [CDF Collaboration],
  Phys.\ Rev.\ Lett.\  {\bf 97}, 171802 (2006)
  [arXiv:hep-ex/0605101];
[D0 Collaboration]
D0 CONF 4400 v1.4
http://www-d0.fnal.gov/Run2Physics/WWW/results/prelim/NP/N06/N06.pdf

\bibitem{Georgi:1992dw}
  H.~Georgi,
  Phys.\ Lett.\  B {\bf 298}, 187 (1993)
  [arXiv:hep-ph/9207278].
  
\bibitem{Georgi:2000ks}
  H.~Georgi, A.~K.~Grant and G.~Hailu,
  Phys.\ Lett.\  B {\bf 506}, 207 (2001)
  [arXiv:hep-ph/0012379].

\bibitem{Chacko:1999hg}
  Z.~Chacko, M.~A.~Luty and E.~Ponton,
  JHEP {\bf 0007}, 036 (2000)
  [arXiv:hep-ph/9909248].
  
 




\end{thebibliography}
\end{document}